\documentclass[sigconf]{acmart}
\usepackage{graphicx}
\usepackage{subcaption}
\usepackage{algorithm}
\usepackage{algorithmic}
\usepackage{siunitx}
\usepackage{float}
\usepackage{bbm}
\usepackage{balance} 
\usepackage{enumitem}
\usepackage{algorithm}

\AtBeginDocument{%
  }

\begin{document}

\title[PRECTR-V2]{PRECTR-V2: Unified Relevance–CTR Framework with Cross-User Preference Mining, Exposure Bias Correction, and LLM-Distilled Encoder Optimization}

\author{Shuzhi Cao}
\affiliation{
  \institution{Alibaba Group}
  \city{Hangzhou}
  \country{China}}
\email{caoshuzhi.csz@taobao.com}
\authornote{All these authors contributed equally to this research.}

\author{Rong Chen}
\affiliation{
  \institution{Alibaba Group}
  \city{Hangzhou}
  \country{China}}
\email{xingshu.cr@taobao.com}
\authornotemark[1]

\author{Ailong He}
\affiliation{
  \institution{Alibaba Group}
  \city{Hangzhou}
  \country{China}}
\email{along.hal@taobao.com}
\authornotemark[1]

\author{Shuguang Han}
\affiliation{
  \institution{Alibaba Group}
  \city{Hangzhou}
  \country{China}}
\email{shuguang.sh@taobao.com}
\authornote{Corresponding Author}

\author{Jufeng Chen}
\affiliation{
  \institution{Alibaba Group}
  \city{Hangzhou}
  \country{China}}
\email{jufeng.cjf@taobao.com}

\renewcommand{\shortauthors}{Shuzhi Cao, Rong Chen, Ailong He, Shuguang Han, \& Jufeng Chen}

\begin{abstract}
In search systems, effectively coordinating the two core objectives of search relevance matching and click-through rate (CTR) prediction is crucial for discovering users' interests and enhancing platform revenue. 
In our prior work PRECTR \cite{chen2025prectr}, we proposed a unified framework to integrate these two subtasks, thereby eliminating their inconsistency and leading to mutual benefit. 
However, our previous work still faces three main challenges. 
First, low-active users and new users have limited search behavioral data, making it difficult to achieve effective personalized relevance preference modeling.
Second, training data for ranking models predominantly come from high-relevance exposures, creating a distribution mismatch with the broader candidate space in coarse-ranking, leading to generalization bias.
Third, due to the latency constraint, the original model employs an Emb+MLP architecture with a frozen BERT encoder, which prevents joint optimization and creates misalignment between representation learning and CTR fine-tuning.
To solve these issues, we further reinforce our method and propose PRECTR-V2. Specifically, we mitigate the low-activity users' sparse behavior problem 
by mining global relevance preferences under the specific query, which facilitates effective personalized relevance modeling for cold-start scenarios.
Subsequently, we construct hard negative samples through embedding noise injection and relevance label reconstruction, and optimize their relative ranking against positive samples via pairwise loss, thereby correcting exposure bias.
Finally, we pretrain a lightweight transformer-based encoder via knowledge distillation from LLM and SFT on the text relevance classification task. This encoder replaces the frozen BERT module, enabling better adaptation to CTR fine-tuning and advancing beyond the traditional Emb+MLP paradigm.
Extensive experimental results demonstrate the effectiveness and superiority of our proposed PRECTR-V2 method.
\end{abstract}

\ccsdesc[500]{Information systems~Data mining}

\keywords{Search System, Click-through Rate Prediction, Exposure Debias, Knowledge Distillation}

\maketitle

\section{Introduction}
Modern search systems are tasked with simultaneously ensuring search relevance and maximizing product conversion efficiency.
While search relevance matching ensures that the items retrieved align with user queries, click-through rate (CTR) prediction captures user interest and thus improves the item's conversion efficiency, driving platform revenue. 
Traditionally, these two objectives have been addressed through separate or loosely coupled models, resulting in inconsistencies in optimization, suboptimal performance, and limited ability to balance user satisfaction with business goals.
In response, we previously introduced PRECTR \cite{chen2025prectr}, a unified and synergistic framework designed to jointly model search relevance and CTR prediction within a single architecture. By aligning the learning signals of both tasks, PRECTR reduced inconsistencies and enabled mutual reinforcement between search relevance matching and CTR prediction. 

Despite these improvements, three fundamental challenges remained unresolved. (1) First, the sparse and fragmented nature of search behavioral data from low-activity and new users severely hindered the learning of personalized relevance preferences, resulting in ineffective user profiling under cold-start scenarios \cite{wei2021contrastive,zhang2025cold}. (2) Second, the training data is subject to significant exposure bias, as models are primarily optimized on historically exposed items with high relevance labels. This created a distributional mismatch between the training environment and the full candidate space encountered during coarse-ranking, ultimately impairing the model's generalization capability. 
(3) Third, due to stringent latency constraints in production environments, the original framework adhered to an Emb+MLP architecture with a frozen pre-trained BERT \cite{koroteev2021bert} text encoder. This design prevents end-to-end joint optimization, leading to a persistent misalignment between the static semantic representations and the dynamic fine-tuning signals derived from click-through behavior.

To address these limitations, we propose PRECTR-V2, an enhanced framework built upon our prior work with three key technical contributions. 
First, to overcome the sparse search behavioral data of low-activity users, we develop a cross-user relevance preference mining framework that extracts and transfers query-conditioned relevance preference patterns from the global user population, enabling effective personalization in cold-start scenarios.
Second, to correct the exposure bias arising from training on predominantly high-relevance items, we construct fake hard negative samples through embedding noise injection and relevance label reconstruction. These samples are then optimized against positive instances via a pairwise ranking loss, explicitly calibrating their relative order and thereby reducing the distribution mismatch between exposure data and the broader candidate space encountered during inference.
Meanwhile, to ensure that the pairwise ranking optimization does not compromise the model's absolute prediction accuracy—measured by PCOC (Predict Click Over Click)—we further introduce two regularization mechanisms: a relative order critical distance penalty and a dynamic truncation weighting strategy. These components work synergistically to balance ranking quality with calibrated CTR estimation.
Third, we introduce a lightweight transformer-based encoder that is pretrained through knowledge distillation \cite{gou2021knowledge,tian2025knowledge,hu2025large} from a large language model (LLM) \cite{naveed2025comprehensive,xi2025rise,guo2024large} and further fine-tuned on a query-item text relevance classification task. This trainable encoder replaces the frozen BERT component in the conventional Emb+MLP pipeline, enabling joint optimization of text representation learning and CTR prediction, and ultimately bridging the gap between semantic understanding and CTR prediction.

The main contributions of this paper are summarized as follows:
\begin{itemize}
  \item We present PRECTR-V2, an improved iteration of our previous PRECTR framework, which addresses three fundamental shortcomings in joint relevance-CTR modeling: (1) ineffective relevance personalization under sparse user interaction, (2) training-inference discrepancy due to exposure bias, and (3) architectural misalignment between semantic representation and CTR prediction under low-latency constraints. 
  \item We introduce three core technical innovations to solve the above issues: (1) A cross-user relevance preference mining mechanism that transfers query-conditioned relevance preference patterns from the global user base to cold-start users, enabling personalized modeling despite sparse individual search behavior.
  (2) An exposure bias correction strategy based on synthetic hard negative sampling via embedding noise injection and relevance label reconstruction, optimized through a pairwise ranking loss regularized by a critical distance penalty and dynamic truncation weighting to preserve CTR calibration.
  (3) A lightweight transformer-based encoder pretrained via LLM distillation and relevance-aware fine-tuning, replacing the frozen BERT module to support joint representation learning and CTR optimization within an efficient inference pipeline.
  \item An extensive amount of experimental analysis on both industrial datasets and online A/B testing demonstrates the effectiveness and superiority of the proposed PRECTR-V2 method over existing methods.
\end{itemize}

The remainder of this paper is organized as follows:
Section \ref{Related work} reviews related work.
Section \ref{PRELIMINARIES} describes the preliminary concepts. 
Section \ref{METHODOLOGY} details the proposed method.
Section \ref{EXPERIMENTS} presents the experimental results.
Finally, Section \ref{conclusion}  concludes this paper.

\section{Related Work}
\label{Related work}
This section provides a concise overview of CTR prediction, search relevance matching, and limitations of prior work.

\subsection{CTR Prediction and Search Relevance Matching}
CTR prediction estimates the likelihood of user clicks on presented items, serving as a core component in advertising, recommendation, and search systems. It directly enhances user engagement, drives revenue growth, and optimizes service efficiency.
The methodology of CTR prediction has progressively transitioned from shallow machine learning-based models to deep learning‑based approaches over time.
Early machine-learning approaches—including logistic regression (LR) \cite{lr}, gradient boosting decision trees (GBDT) \cite{gbdt}, and factorization machines (FM) \cite{fm}—primarily modeled user–item co‑occurrence patterns and depended heavily on hand‑crafted feature engineering.
Subsequent deep learning‑based methods, including DeepFM \cite{guo2017deepfm}, DIN \cite{DIN}, DIEN \cite{DIEN}, SIM \cite{SIM}, and TWIN \cite{chang2023twin}, learn continuous, dense embedding representations and model CTR predictions in an end‑to‑end manner.
Recently, fueled by the widespread success and scaling capabilities of large language models (LLMs), both industry and academia have begun to explore the autoregressive generation paradigm of LLMs and adapt it to search and recommendation scenarios. This approach reformulates the traditional discriminative CTR prediction task as a long‑sequence generation problem, where candidate product representations are directly generated in an autoregressive manner \cite{yu2025randomized,yao2025denoising}.
Among them, representative works include OneSearch \cite{onesearch}, OneRec \cite{zhou2025onerec}, GenRank \cite{genrank}, and LUM \cite{lum}.
Nevertheless, existing methods primarily address click‑through rate prediction in isolation, neglecting its inherent coupling with search relevance matching and thereby introducing a misalignment between the two fundamental tasks.

A search system must balance two core objectives: maximizing click‑through rates while preserving strong relevance to user intent—both of which ultimately drive conversion.
In conventional search pipelines, CTR prediction and relevance scoring are handled by separate, independently trained models. Their outputs are then combined using a fixed, human‑designed strategy to compute the final ranking score. While this decoupled design is straightforward to implement, it limits feature interaction between the two tasks and introduces inconsistency across their predictions, ultimately leading to suboptimal system performance.
To address this challenge, we propose PRECTR \cite{chen2025prectr}—the first unified framework that simultaneously handles personalized search relevance matching and CTR prediction. By integrating both tasks into a single model, we effectively mitigate the inconsistency between them. Through joint optimization, the framework automatically learns the optimal fusion strategy, eliminating the need for laborious manual parameter tuning. This approach not only enhances adaptability to shifting data distributions but also improves overall recommendation robustness and efficiency.

\subsection{PRECTR's Limitations}

Despite its unified architecture that jointly models search relevance matching and CTR prediction, PRECTR exhibits several critical limitations that constrain its practical effectiveness. These shortcomings primarily emerge from three key aspects: (1) personalization challenith for low-active users (2) exposure bias in training and (3) architectural constraints due to latency requirements.

First, PRECTR's personalized relevance modeling struggles with low-activity and new users whose search interactions are sparse or non-existent. This data scarcity prevents the framework from constructing meaningful user relevance preference representations, leading to ineffective personalization in cold-start scenarios.
Second, PRECTR suffers from significant exposure bias as it is predominantly trained on historically exposed items with high relevance degree. This creates a distributional mismatch between the training environment and the actual inference scenario, where the model must evaluate a broader candidate pool. The resulting overfitting to high-relevance exposures compromises the model's generalization capability, reducing ranking diversity and limiting the system's ability to discover novel relevant matches. 
Third, to meet production latency requirements, PRECTR adopts an Emb+MLP architecture with a frozen pre-trained BERT encoder. This design prevents joint optimization of text representation learning and CTR prediction, creating a persistent misalignment between static semantic features and dynamic engagement signals. The frozen encoder cannot adapt to domain-specific nuances or evolving user behavior patterns, while the shallow MLP limits the model's capacity to capture complex feature interactions essential for both relevance understanding and click prediction.

\section{Preliminaries}
\label{PRELIMINARIES}

\subsection{Problem Formulation}
Given the training dataset $\mathcal{D}=\left\{\boldsymbol{x},y\right\}^{n}$ collecting from our online e-commerce platform Xianyu, where $\boldsymbol{x}$ represents the high dimensional feature vector consisting of multi-fields features, the binary click label $y \in \left\{0,1\right\}$ indicates whether a sample is clicked, and $n$ denotes the number of training samples.
The feature $\boldsymbol{x}$ usually consists of multi-fields information and can be denoted as $\boldsymbol{x}=[t_{1}^\top,t_{2}^\top,...,t_{M}^\top]^\top$ where $t_{i} \in \mathbb{R}^{K_{i}}$ stands for the high-dimensional sparse binary features for $i$-th field, $K_{i}$ represents the dimensionality of vector $t_{i}$, and $M$ stands for the number of fields.
$t_{i}$ can be either a one-hot vector or a multi-hot vector depending on the number of values of the $i$-th field.
CTR prediction task aims to estimate the probability of a sample $x$ being clicked,i.e., $p_{ctr}(x)=p(y=1|x)$.
Similarly, the search relevance matching task can be formulated as below: given the original query text $Q_{text}$ and item description text $I_{text}$, our goal is to determine whether the two texts are relevant to each other.
In Xianyu, for each user query, all of the candidate items are divided into four different relevance score levels based on their relevance, i.e., the degree of textual matchness between the query and the item.

\subsection{The Review of PRECTR}
In PRECTR, we decompose the probability of clicking $P(click=1|\boldsymbol{x})$ on an item $\boldsymbol{x}$ into two parts: the relevance to the user query and the intrinsic item characteristic. 
To be specific, we indivdually model $P(rsl=i|\boldsymbol{x})$ (the probability of the distribution of relevance score level (RSL) under the given instance $\boldsymbol{x}$) and $P(click=1|rsl=i,\boldsymbol{x})$ (the 
conditional probability of the item $\boldsymbol{x}$ being clicked in the $i$-th relevance score level) and combine them to estimate the click rate $P(click=1|\boldsymbol{x})$.
Finally, by analyzing users' personalized relevance preferences derived from their sequential search behavior, we compute the relevance incentive score $\tau(\boldsymbol{x})$ to refine the originally estimated click-through rates $P(click=1|\boldsymbol{x})$ and obtain the final ranking score as follows:

\begin{equation}
\label{eq1}
P(click=1|\boldsymbol{x})=\sum_{i=1}^{k}P(click=1|rsl=i,\boldsymbol{x})P(rsl=i|\boldsymbol{x}),
\end{equation}

\begin{equation}
\label{eq2}
rank\_score=\tau(\boldsymbol{x})*P(click=1|\boldsymbol{x}),
\end{equation}
Here, the variable $rsl$ stands for the relevance score level, which can be any of the $k$ discrete values.

\section{Methodology}
\label{METHODOLOGY}
\begin{figure*}
    \centering
    \includegraphics[width=0.9\hsize,height=0.56\hsize]{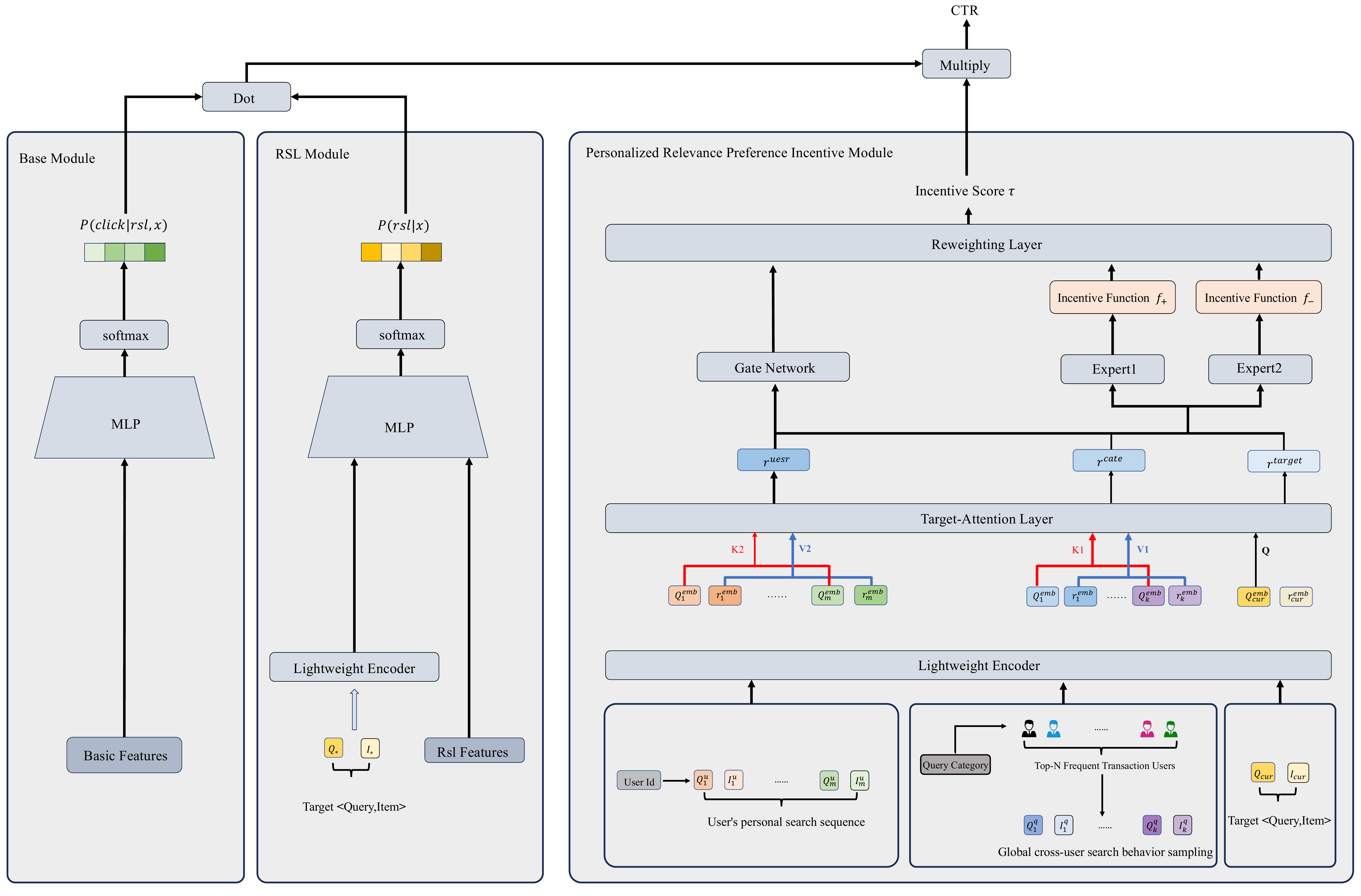}
    \caption{The overview of the proposed PRECTR-V2 method.}
    \label{overview}
\end{figure*}

This section outlines the overall design of our proposed method, including the architecture overview and the detailed structure.
Briefly, PRECTR-V2 includes three innovative parts: (1) Cold‑Start Personalized Relevance Preferences Mining, (2) Relevance Exposure Debias, and (3) LLM-Distilled CRT-Aligned Lightweight Encoder. The model's whole structure is shown in Fig.\ref{overview}

\subsection{Cold‑Start Personalized Relevance Preferences Mining}
For users with rich interaction histories,
we extract the users' personal relevance preferences from their historical interaction records $\mathcal{S}=\left\{(Q_{1}, I_{1}),...,(Q_{m}, I_{m})\right\}$,
where $m$ denotes the length of the historical click sequence, $Q_{i}$ and $I_{i}$ represent the raw text of the query and clicked item, respectively.
However, for low-activity and new users, their sparse search behaviors pose significant challenges for modeling relevance preferences.
To mitigate this limitation, we enhance the representation of low-activity users by leveraging global information from cross-user behavior patterns, supplementing their sparse individual search data.
Data analysis reveals a close association between users' sensitivity to search relevance and the category of their queries.
The average relevance level of clicked samples varies significantly across different search query categories, suggesting that query type is a key determinant of perceived relevance.
Inspired by this phenomenon, we leverage the retrieval of global user behaviors within the same query category to extract common preference patterns.
Specifically, based on the user’s query category, we globally sample 50 users who have exhibited similar categorical search behavior and extract their corresponding historical search records.
Subsequently, we retrieved the historical search click behaviors within the same category from these sampled users, filtered the top-k behaviors based on query text similarity, and used them as supplementary behavior sequences for low-activity users.
Ultimately, we augmented the sparse search click behaviors of low-activity users by integrating cross-user sampled behaviors, thereby expanding their effective search interaction information.
We use the $\mathcal{S}=\left\{\mathcal{S}_{u},\mathcal{S}_{q}\right\}$ to denote the merged search sequence of the cold-start users, where $\mathcal{S}_{u}=\left\{(Q_{1}^{u},I_{1}^{u}),...,(Q_{m}^{u},I_{m}^{u})\right\}$ and $\mathcal{S}_{q}=\left\{(Q_{1}^{q},I_{1}^{q}),...,(Q_{k}^{q},I_{k}^{q})\right\}$ represent the low-active users' own and cross-users' search behaviors, where $m$ and $k$ stands for the length of these two sequences.
We first take the constructed text sequence as input and leverage a language encoder to derive its textual representations, which comprise the following query, item, and relevance embeddings.
Their definitions are shown as follows:
\begin{equation}
\label{eq3}
Q_{i}^{emb} = \text{Encoder}(Q_{i})
\end{equation}
\begin{equation}
\label{eq4}
I_{i}^{emb} = \text{Encoder}(I_{i})
\end{equation}
\begin{equation}
\label{eq5}
r_{i}^{emb} = \text{Encoder}(\text{[CLS]}Q_{i}\text{[SEP]}I_{i}\text{[SEP]})
\end{equation}

Let $Q_{cur}$ denote the raw text of the user's current query and $Q_{cur}^{emb}$ represent its embedding after the language model encoding.
We use $Q_{u}^{emb}=[Q_{1}^{emb},..., Q_{m}^{emb}]$,
$r_{q}^{emb}=[r_{1}^{emb},..., r_{m}^{emb}]$, 
$Q_{q}^{emb}=[Q_{1}^{emb},..., Q_{k}^{emb}]$, and $r_{u}^{emb}=[r_{1}^{emb},..., r_{k}^{emb}]$ to represent the embeddings of these two sequences.
Subsequently, we estimate the search relevance preference representations at the user's level and query category's level, representing as $r_{u}^{emb}$ and $r_{q}^{emb}$ through the Multi-Head Target-Attention (MHTA) operation.
Their definitions are shown as follows:
\begin{equation}
\label{eq6}
    r^{user}=\text{TargetAttention}(Q,K_{u},V_{u})=softmax(\frac{Q  K_{u}^\top}{\sqrt{d}})V_{u},
\end{equation}
\begin{equation}
\label{eq7}
    r^{cate}=\text{TargetAttention}(Q,K_{q},V_{q})=softmax(\frac{Q  K_{q}^\top}{\sqrt{d}})V_{q},
\end{equation}
where $Q$ is derived from $Q_{\text{cur}}$; $K_u$ and $K_q$ originate from $Q_u$ and $Q_q$, respectively; and $V_u$ and $V_q$ are obtained from $r_u$ and $r_q$, respectively. The temperature factor $\sqrt{d}$ is applied to produce a smoother attention distribution, thereby preventing excessively small gradients during training. 
After estimating the search relevance preference representations at both the user level and the query category level, we further introduce a Mixture of Experts (MoE) network to compute the final personalized relevance incentive score $\tau$.
Specifically, we feed $r^{user}$ into a gating network to generate a two-dimensional soft label distribution $w = [w_{1}, w_{2}]$, which reflects the user's relevance preference and sensitivity and is used to weight the expert networks.
Correspondingly, we design two distinct expert networks, each employing a different activation function, to process users with varying relevance preferences. The input to each expert includes $r^{user}$, $r^{cate}$, and the target relevance embedding $r^{cur}$. The two activation functions are formulated as follows:
\begin{equation}
\label{eq7}
    f_{+}(\boldsymbol{x})=\text{log}(1+e^{E_{+}(\boldsymbol{x})})
\end{equation}
\begin{equation}
\label{eq8}
    f_{-}(\boldsymbol{x})=-\text{log}(1+e^{E_{-}(\boldsymbol{x})})
\end{equation}
where $E_{+}$ and $E_{-}$ stands for two different expert networks respectively.
Finally, the incentive score can be calculated by combining weighted outputs as follows:
\begin{equation}
\label{eq9}
    \tau = w1*f_{+}(\boldsymbol{x})+w2*f_{-}(\boldsymbol{x})
\end{equation}

\subsection{Relevance Exposure Debias}
The CTR model is trained on real exposure data. Due to the stringent relevance constraints applied during exposure, the training samples predominantly exhibit highly correlated features—our analysis indicates that over 80\% of exposed samples fall into this category. However, the actual inference environment of the fine-grained model involves scoring coarsely filtered candidate items, which consist largely of products with low search relevance. This significant distribution shift between the training and inference data severely compromises the model's generalization ability.
This mismatch motivates our work in addressing the resulting bias.

Intuitively, we can supplement the training data by sampling irrelevant instances from the unexposed candidate pool to guide the model learning process.
However, this straightforward method severely influence the ratio of positive and negative training samples. 
Meanwhile, unexposed unrelated samples are often of low quality—for instance, they may be entirely irrelevant or possess extremely poor features on the item side. This makes it difficult for the model to discern scoring variations attributable to relevance-induced differences.
Therefore, we propose to mitigate the learning bias by constructing synthetic low-relevance samples while keeping other item-side features aligned through embedding noise injection and relevance label reconstruction.
Suppose $rsl$ denotes the relevance score level of an item, which takes four discrete values: 1, 2, 3 and 4, representing irrelevant, weak relevant, relevant, and strongly relevant, respectively.
We construct negative samples $\boldsymbol{x}{-}$ in a 1:1 ratio for those positive samples $\boldsymbol{x}{+}$ that are both clicked and highly relevant (i.e., $click=1$, $rsl=4$).
Specifically, for each such positive sample, we reset its corresponding $rsl$ label to $fake\_rsl$ by randomly assigning it to a lower relevance level according to the following scheme, where $p1$ and $p2$ are presetting hyperparameters.
\begin{equation}
\label{eq10}
    fake\_rsl = \begin{cases}
    1 & \text{if } random()<p1 \\
    2 & \text{if }  p1 \leq random()<p2 \\
    3 & \text{if } p2 \leq random()<1
\end{cases}
\end{equation}

Subsequently, we simulate slight variations in search relevance matching degree by adding noise to the textual representation of item descriptions.
The noise injection process is shown in Eq.(\ref{ee}).
Let $Q_{emb} \in \mathbb{R}^{d}$ be the original 
d-dimensional text embedding vector. We generate a noise vector $\epsilon \in \mathbb{R}^{d}$ by independently sampling each component $\epsilon_{i}$ 
from a standard normal distribution $\mathcal{N}(0, 1)$. The perturbed embedding $Q_{\mathrm{emb}}^{\mathrm{fake}}$ is then obtained by performing element-wise addition between the original embedding and the noise vector. 

\begin{equation}
\label{ee}
    Q_{\mathrm{emb}}^{\mathrm{fake}} = Q_{\mathrm{emb}} + \epsilon
\end{equation}

\begin{equation}
        \epsilon = \bigl( \epsilon_1, \epsilon_2, \dots, \epsilon_d \bigr)^\top,
    \epsilon_i \stackrel{\mathrm{i.i.d.}}{\sim} \mathcal{N}(0, 1)
\end{equation}

Under the premise that other features are aligned, $\boldsymbol{x}_{+}$ and its corresponding constructed negative sample $\boldsymbol{x}_{-}$ differ only in the relevance features.
Subsequently, we conduct relevance exposure debias by explicitly imposing a constraint on the relative order between $\boldsymbol{x}_{+}$ and $\boldsymbol{x}_{-}$ through a pairwise loss function as follows:
\begin{equation}
    \label{eq12}
    loss = \sum_{i=1}^{N}\text{log}(1+e^{-(f(x_{i}^{+})-f(x_{i}^{-}))})
\end{equation}
where $N$ stands for the number of the positive samples, $f$ denotes the whole rank model. By optimizing the Eq(\ref{eq12}), we force the model to assign higher scores to positive samples, which in turn enhances its ability to discern relevance differences. In practice, however, we have observed that the aforementioned loss functions suffer from the following limitations.

\begin{figure}
    \centering
    \includegraphics[width=0.95\hsize,height=0.7\hsize]{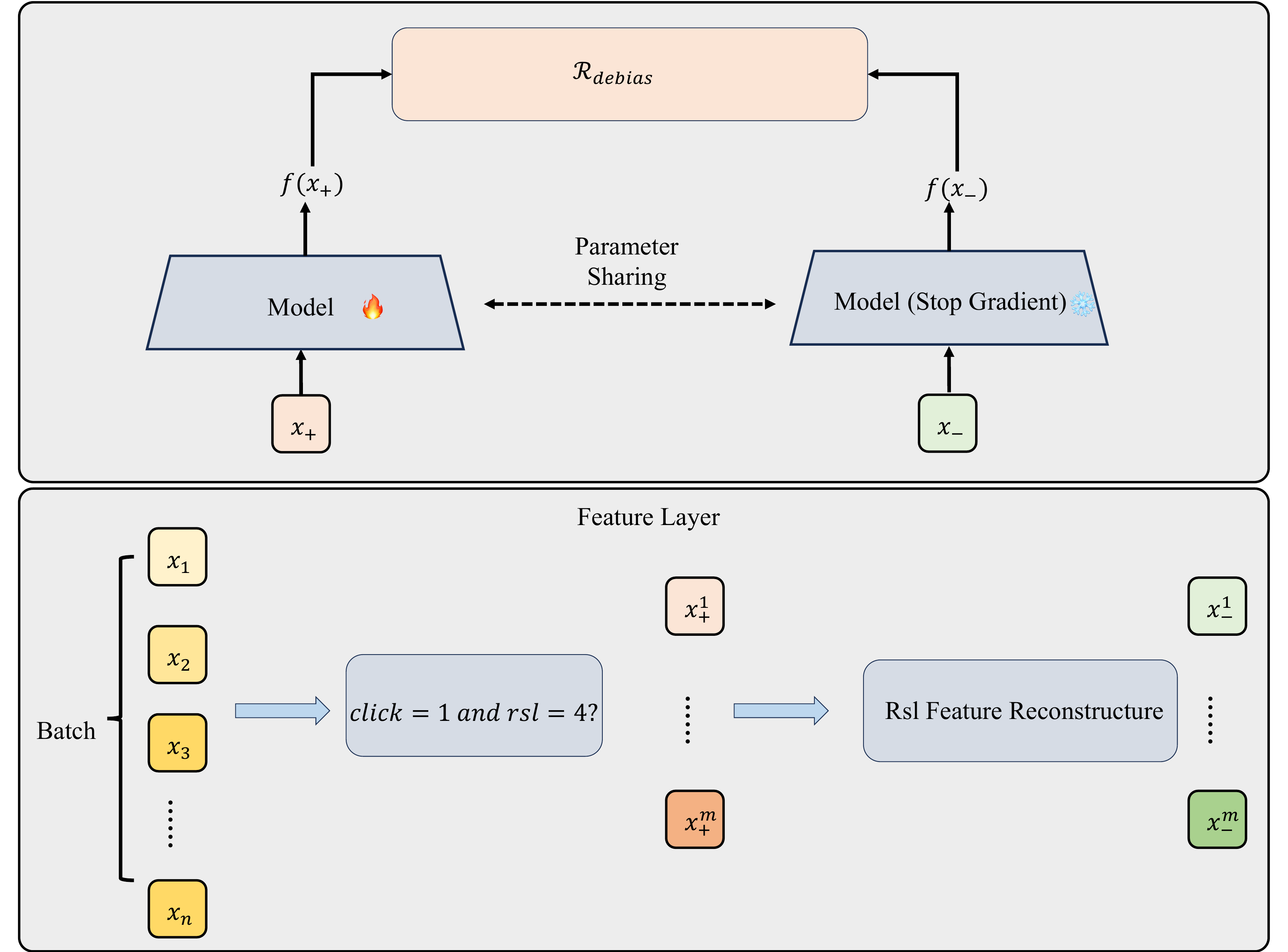}
    \caption{The overview of the debias process.}
    \label{debias}
\end{figure}

\begin{figure}[h]
	\centering
	\begin{subfigure}{0.495\linewidth}
		\centering
		\includegraphics[width=1\linewidth]{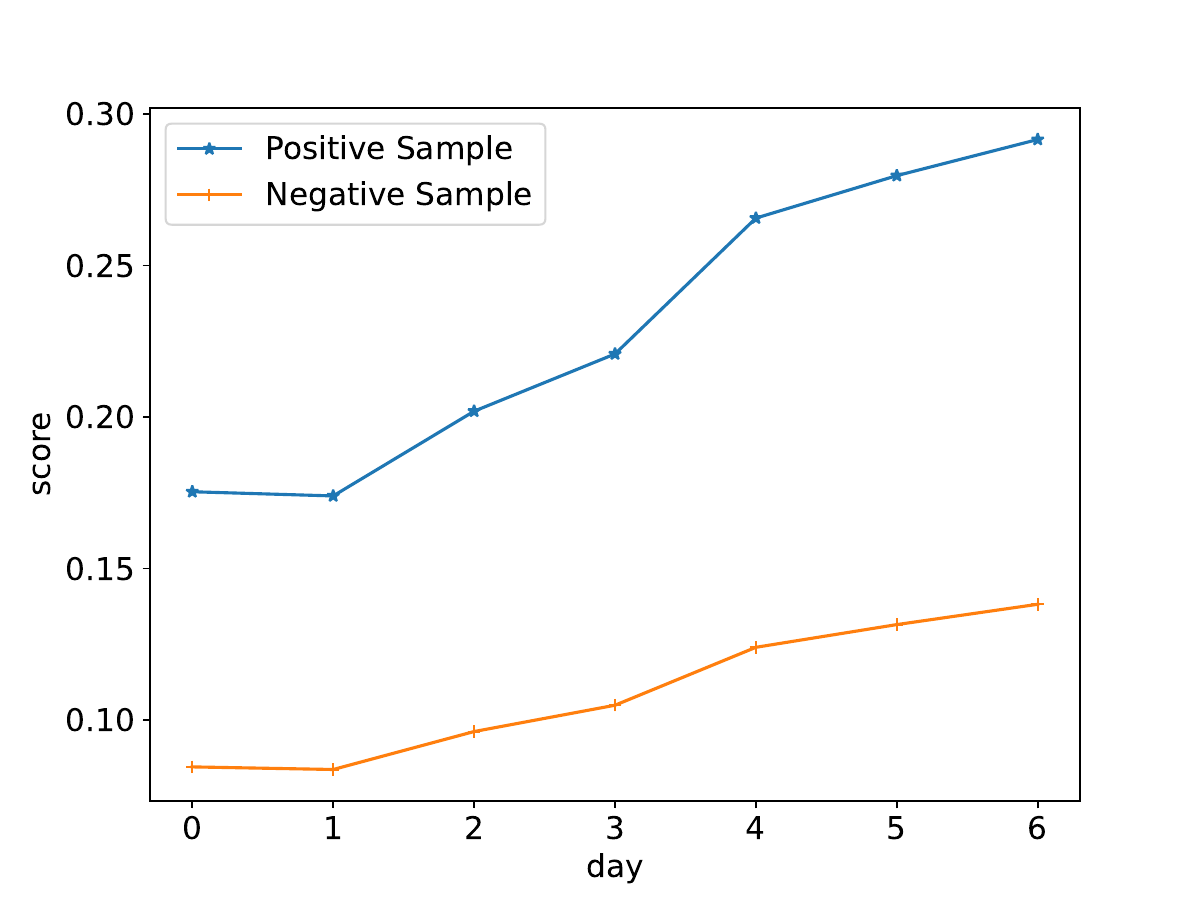}
		\caption{}
		\label{pie}
	\end{subfigure}
	\centering
	\begin{subfigure}{0.495\linewidth}
		\centering
        \includegraphics[width=1\linewidth]{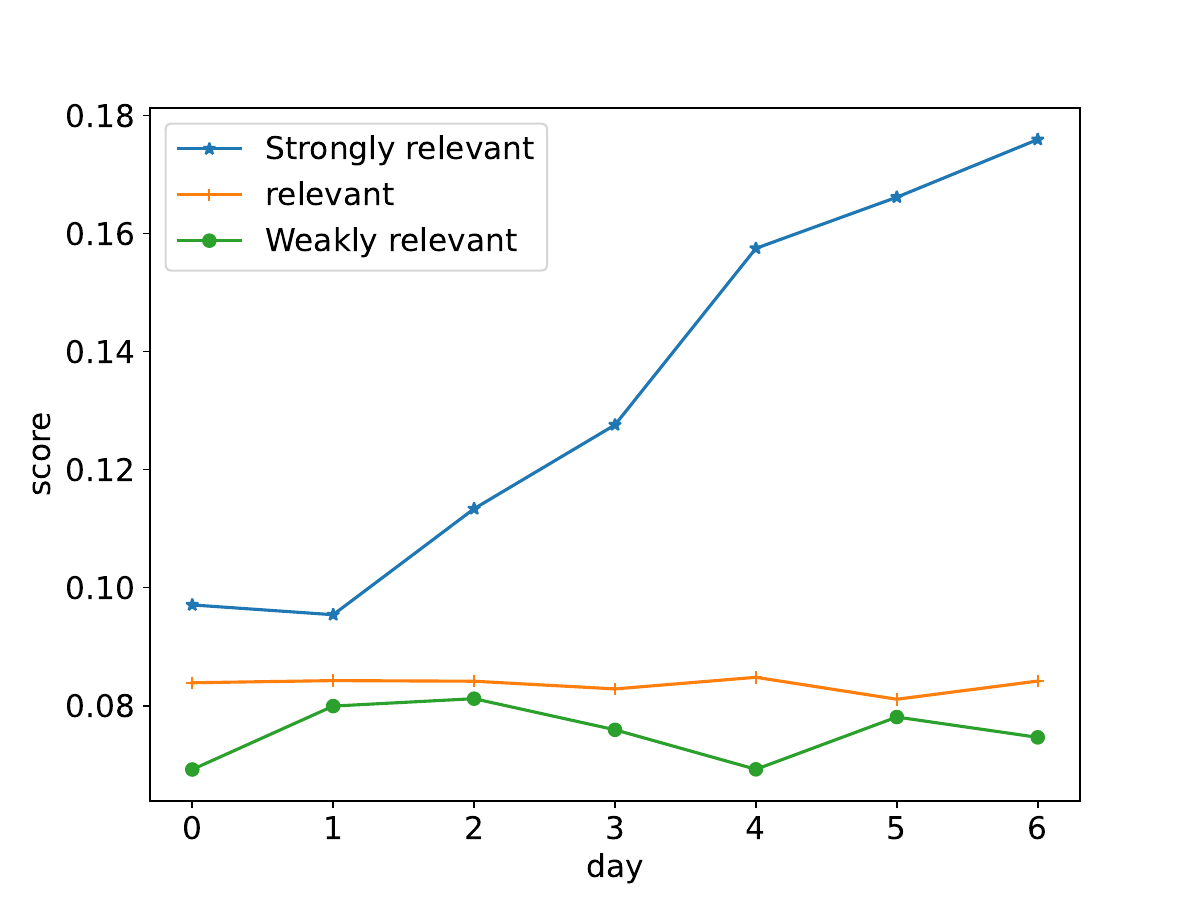}
        \caption{}
		\label{t-sne}
	\end{subfigure}
	\caption{(a) exhibits the predicted average score between the click and unclick samples, and (b) exhibits the predicted average score difference between the samples with different relevance score levels.}
	\label{fig1new}
\end{figure}

The Fig.\ref{fig1new} illustrates that sample scores exhibit growing divergence with increased training time.
Specifically, the average score of clicked samples monotonically increases, while the performance gap observed between clicked and unclicked samples continues to widen.
Meanwhile, the predicted scores of high-relevance samples have been continuously rising, even reaching more than twice that of low-relevance samples.
All these factors severely degrade the absolute scoring accuracy of the model, 
causing the deviation of PCOC (Predict Click Over Click).
Moreover, it is also unreasonable to infinitely magnify the relative order between 
$\boldsymbol{x}_{+}$ and $\boldsymbol{x}_{-}$.
To solve these problems, we further propose a relative order critical distance penalty and a dynamic truncation weighting strategy.
These components work synergistically to balance ranking quality with calibrated CTR estimation.
The refined loss is shown as follows:
\begin{equation}
    \label{eq13}
    \mathcal{R}_{debias} = w(x)* \sum_{i=1}^{N}\text{log}(1+e^{\text{max}(0,margin-(f(x_{i}^{+})-f(x_{i}^{-}))})
\end{equation}
where $\mathit{margin}$ is the penalty threshold for the relative order. It denotes the maximum permissible difference in scores between $\boldsymbol{x}{+}$ and $\boldsymbol{x}{-}$.
$w(x)$ is a dynamic truncation weight coefficient, it dynamically adjusts the loss weight based on the average score of a batch of samples. If the batch average score exceeds a predefined critical threshold, the control weight is reduced to zero, effectively applying dynamic truncation. Its definition is shown below:

\begin{equation}
\label{eqthres}
    w(x) = \begin{cases}
    w & \text{if } \text{mean}(f(x_{+})) <threshold \\
    0 & \text{if } \text{mean}(f(x_{+})) \geq threshold \\
\end{cases}
\end{equation}

\subsection{LLM-Distilled CTR-Aligned Lightweight Encoder.}
In PRECTR \cite{chen2025prectr}, we employ a frozen BERT model as the base encoder for text embedding generation, followed by an Emb+MLP architecture for downstream processing. This design, however, comes with a trade-off: the base encoder is excluded from CTR-specific fine-tuning. The primary reason is its prohibitive parameter scale, which poses significant challenges for efficient online deployment and inference.
To overcome this drawback, we propose an LLM-distilled and CTR-aligned lightweight encoder to replace the previous frozen BERT.
We construct a lightweight encoder consisting of three stacked Transformer layers. With only about 2M parameters, it is extremely compact compared to the BERT-base model (110M).
This lightweight encoder is initialized via supervised fine-tuning (SFT) \cite{chu2025sft} for relevance classification and embedding distillation from a large model, and then jointly fine-tune it end-to-end with the CTR prediction task.
The following will introduce its details.

\textbf{Textual Relevance Classification SFT.} In this stage, we sample <query, item> text pairs with different relevance score levels ($rsl$) in the exposure space and use their relevance label as supervisory signals to conduct the text classification task.
Specifically, suppose $T(\boldsymbol{x};\theta)$ represents the semantic embedding of the text pair generate by the encoder $T$ parameterized by $\theta$.
We use an additional projection network $M$ to convert its to the relevance label space, i.e, $M(T(\boldsymbol{x},\theta))=[P(rsl=1|\boldsymbol{x}),..., P(rsl=4|\boldsymbol{x})]^\top$.
Then we use $rsl$ as the supervised label, and the SFT risk is in cross-entropy \cite{connor2024correlations} form:
\begin{equation}
\label{eq15}
    \mathcal{R}_{SFT}(\theta)=-\frac{1}{n}\sum_{i=1}^{n}rsl*\text{log}(softmax(M(T(\boldsymbol{x};\theta)))),
\end{equation}

\begin{figure}
    \centering
    \includegraphics[width=0.95\hsize,height=0.54\hsize]{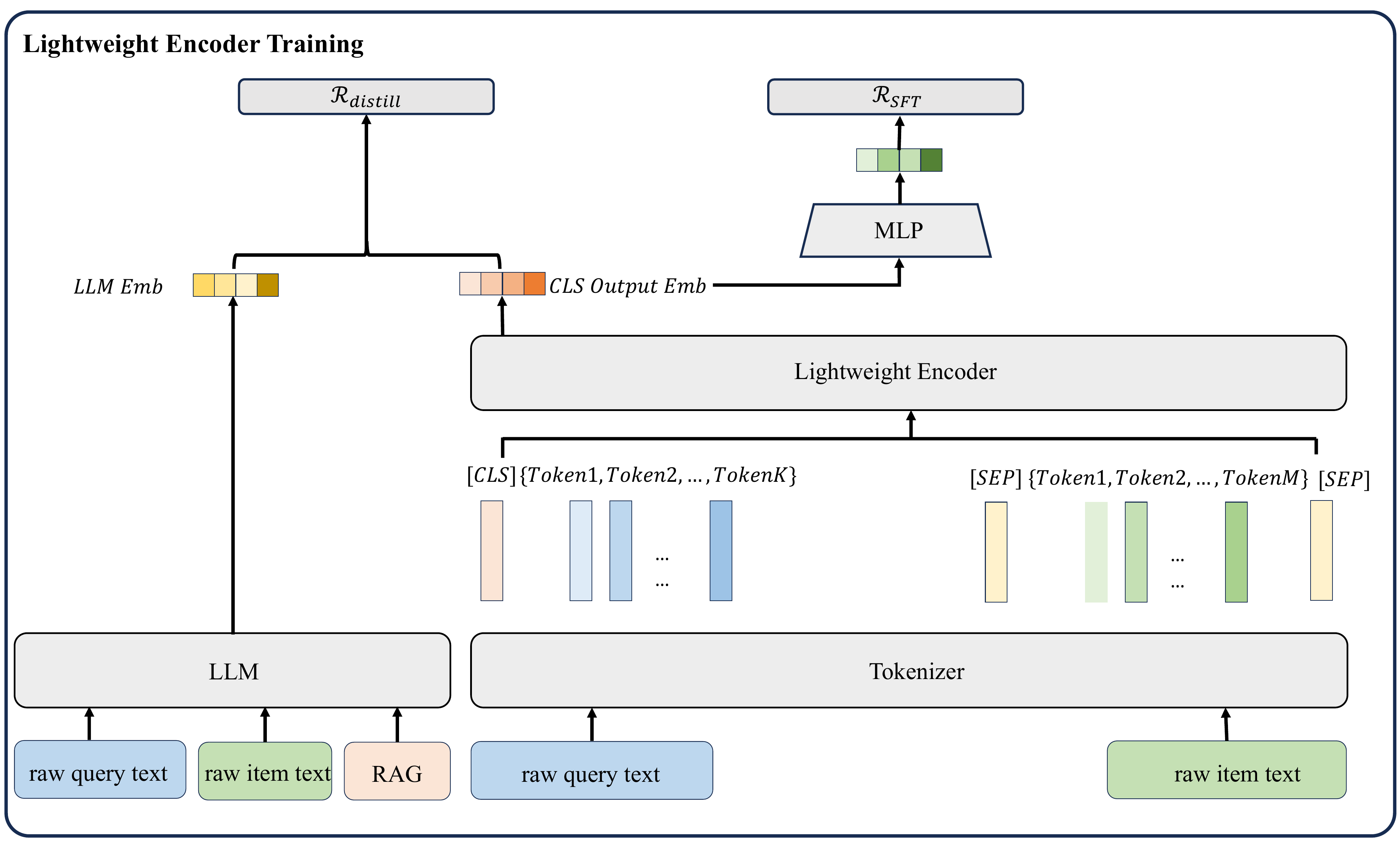}
    \caption{The overview of the lightweight encoder training.}
    \label{leightweight}
\end{figure}

\textbf{LLM Embedding Distillation.}
We adopt knowledge distillation for model compression, where Qwen-7B acts as the teacher to transfer knowledge to our lightweight student encoder.
This process enables our model to acquire the open-world knowledge inherent in LLMs and produce higher-quality text representations.
For the training query-item text pairs, we first utilize Qwen-7B \cite{bai2023qwen,yang2025qwen2} to generate their text representations.
In this process, we further introduce the Retrieval-Augmented Generation (RAG) \cite{oche2025systematic,gupta2024comprehensive,siriwardhana2023improving} technique to generate better relevance embeddings.
To be specific, we retrieve highly related samples and relevance badcases under the same query as template information to guide the LLM's inference.
Let $g(\boldsymbol{x}) \in \mathcal{R}^{d}$ be the representation computed by LLM.
We minimized the Euclidean distance between $g(\boldsymbol{x})$ and $T(\boldsymbol{x};\theta)$ and obtained the following MSE risk. The whole lightweight encoder training process is shown in Fig.\ref{leightweight}.
\begin{equation}
\label{eq16}
    \mathcal{R}_{Distill}(\theta)= \text{MSE}(g(\boldsymbol{x}),T(\boldsymbol{x};\theta))
\end{equation}

Finally, we train the lightweight encoder by optimizing $\mathcal{R}_{Distill}(\theta)$  and $\mathcal{R}_{SFT}(\theta)$ simultaneously. The overall optimization objective is concluded as follows:

\begin{equation}
    \mathcal{R}_{overall} = \mathcal{R}_{Distill}(\theta) + \mathcal{R}_{SFT}(\theta)
\end{equation}

\textbf{End-to-End Tuning.}
Following pretraining, we embed the lightweight encoder into our CTR model. Subsequently, the integrated model undergoes end-to-end fine-tuning on user behavior data for final alignment, with a reduced learning rate applied to the encoder parameters. These refinements have collectively enhanced the model's foundational encoding capabilities.

\section{Experiments}
\label{EXPERIMENTS}
In this section, we conduct comprehensive experiments on both the offline dataset and online A/B testing to evaluate the effectiveness and superiority of the proposed method.

\subsection{Experiments Setup}
\textbf{Dataset:} We collect click traffic logs from Alibaba's second-hand online trading platform Xianyu to construct the training dataset. 
Specifically, we treat clicked items as positive samples and unclicked items as negative samples. 
The daily training data volume is about 1.6 billion and each record contains 651 features  (e.g., user, query, and item features).
We take 9 days of data for the experiment evaluation,  in which data from the first 7 days is used for model training, and the remaining data of 2 days are used for model testing.

\textbf{Compared Methods:}  To demonstrate the superiority of the proposed method, we adopt the following baseline approaches:
\begin{itemize}
    \item \textbf{LR \cite{Logiticreg}:} Logistic Regression (short for LR) is a widely used shallow model before the deep model that estimates the CTR by combining features in a linear combination form.
    \item  \textbf{DNN:} DNN is proposed by YouTube and has been widely used in industry scenarios. It follows classic Embedding \& MLP architecture and utilizes the SumPooling operation to integrate historical behavior embeddings.
    \item \textbf{Wide\&Deep \cite{cheng2016wide}:} consists of both wide linear models and deep neural networks to benefit from memorization and generalization mutually. It is selected as our base model.
    \item \textbf{DeepFM \cite{guo2017deepfm}:} It combines factorization machine (FM) and deep neural network to improve Wide\&Deep, effectively capturing both low and high-order feature interactions.
    \item \textbf{XDeepFM \cite{lian2018xdeepfm}:} which proposes a novel Compressed Interaction Network (CIN) to generate feature interactions explicitly and combine it with DNN to predict CTR.
    \item \textbf{DIN \cite{DIN}:} which proposes a novel deep interest network and first utilizes the target attention operation to assess the relevance of the candidate item to previously clicked items for discovering the items that users are interested in.
    \item  \textbf{SuKD \cite{SupplementaryNLPFeatures}:} which utilizes the pre-trained BERT model to encode the raw text of query and item and take it as a supplementary feature to boost CTR prediction.
    \item  \textbf{PRECTR} \cite{chen2025prectr} which first propose a unified framework to integrate the personlized search relevance matching and CTR prediction.

\end{itemize}
\begin{table*}[ht!]
    \centering
    \renewcommand{\arraystretch}{1.15}
    \setlength{\tabcolsep}{29pt}  
    \caption{The offline comparison results, where "RI" is short for "RelaImpr".}
     \begin{tabular}{c | c c |c c }
          \toprule[1.2pt]
          Method               &   AUC      & RI        &  GAUC    &   RI        \\ \hline
          LR \cite{Logiticreg} & 0.6795     &  -29.10\% & 0.6347   & -26.6\%    \\
          DNN                  & 0.7541     & 0.35\%    & 0.6863	  & 1.47\%      \\
          Wide\&Deep \cite{cheng2016wide} &0.7532	&0.00\%  & 0.6836	&0.00\%	\\
          DeepFM \cite{guo2017deepfm} &0.7519 &-0.51\% & 0.6839 &0.16\%       \\
          XDeepFM \cite{lian2018xdeepfm} & 0.7521	&-0.43\%  &0.6843	&-0.29\% \\
          DIN \cite{DIN} &0.7561	&1.15\% &0.6875  &2.21\%  \\
          SuKD \cite{SupplementaryNLPFeatures} &0.7524  &-0.31\% & 0.6851 &0.81\%  \\
          PRECTR \cite{chen2025prectr} & $0.7581$	& $1.93\%$ & $0.6892$ & $3.05\%$  \\
          \hline
          PRECTR-V2   & $\boldsymbol{0.7674}$	& $\boldsymbol{5.61\%}$ & $\boldsymbol{0.6933}$ & $\boldsymbol{5.28\%}$  \\
          \bottomrule[1.2pt]
    \end{tabular}
    \label{table1}
\end{table*}

\textbf{Evaluation Metric:} The definition of the main evaluation metric used in the experiments is presented as follows:
\begin{itemize}[leftmargin=*]
  \item \textbf{AUC:} As the most commonly used evaluation metric in the search recommendation system, Area Under the Curve (AUC) reflects the sorting ability of the CTR model. 
  Specifically, given a positive and a negative item chosen randomly, AUC shows the likelihood that the model would rate the positive item higher than the negative one.
  Therefore, the AUC can be formulated as follows:
  \begin{equation}
      \text{AUC}=\frac{1}{|P||N|}\sum_{p\in P}\sum_{n\in N}\mathbbm{1}(\Theta(p)>\Theta(n)),
  \end{equation}
  where $P$ and $N$ represent the positive and negative item set, respectively,  $\Theta$ is the ranking function given by the CTR model, and $\mathbbm{1}$ is the indicator function.
  \item \textbf{GAUC:} Different from the AUC that measures the global ranking ability of the model, the Group Area Under the Curve (GAUC) is designed to measure the goodness of order by ranking towards various groups or users. In specific, we first calculate AUC for different users and subsequently average them to get the final GAUC.
  The definition of GAUC is formulated as follows, where $\text{AUC}_{i}$ stands for the AUC for the $i$-th user, $\text{\#impression}_{i}$ is its corresponding weight, and $n$ denotes the total number of users.
  \begin{equation}
      \text{GAUC}=\frac{\sum_{i=1}^{n}(\#\text{impression}_{i}\times\text{AUC}_{i})}{
      \sum_{i=1}^{n}(\#\text{impression}_{i})
      }.
  \end{equation}
  \item \textbf{RelaImpr:} RelaImpr is adopted to measure the relative improvement over other models. $\text{RelaImpr}>0$ means the current model is superior over the base model, and vice versa. It can be formulated as follows:
  \begin{equation}
      \text{RelaImpr}=(\frac{\text{AUC}(\text{measured model})-0.5}{\text{AUC}(\text{base model})-0.5}-1)*100\%
  \end{equation}
\end{itemize}

\textbf{Implementation Details:} In all of the experiments, the batch size is set to $4096$, the learning rate is set to 1e-4, and the SGD optimizer is employed for model update.
The history click sequence is collected with the last 30 days, and the maximum length is 25.
The $p1$ and $p2$ in Eq(\ref{eq10}) is set as 0.2 and 0.6 respectively through hyperparameter tuning. The $threshold$ in Eq(\ref{eqthres}) is set as $0.08$ to be 
consistent with the average online click-through rate of the search samples.
The $margin$ in $\mathcal{R}_{debias}$ is set as 0.075, which is equal to the average scoring deviation between positive and negative samples.

\begin{table*}[ht!]
    \centering
    \renewcommand{\arraystretch}{1.2}
     \setlength{\tabcolsep}{19pt}  
    \caption{The ablation study of different PRECTR-V2 variants on production datasets, where "w/o" is short for "without".}
     \begin{tabular}{c| c c | c c}
          \toprule[1.2pt]
          Method            &   AUC   & RI      &  GAUC      & RI  \\ \hline
          PRECTR            & 0.7581  & 0.00\%  &  0.6892    & 0.00\%    \\
          w/o Cold-Start Relevance Preference Mining & 0.7609  & 1.08\%  & 0.6908 & 0.85\%\\
          w/o Relevance Exposure Debias  & 0.7658  &2.98\%  & 0.6928 & 1.91\%\\
          w/o LLM-Distilled and CTR-Aligned Encoder        & 0.7662  & 3.13\%  & 0.6927 & 1.85\%\\
          \hline
          PRECTR-V2  & $\boldsymbol{0.7674}$ & $\boldsymbol{3.60\%}$ & $\boldsymbol{0.6933}$ & $\boldsymbol{2.16\%}$\\
          \bottomrule[1.2pt]
    \end{tabular}
    \label{table2}
\end{table*}

\subsection{Experimental Results}
In this subsection,
We conclude the experimental results from the following five aspects:

\textbf{(1) How is the offline experimental results?} 
The offline experimental results, summarized in Table \ref{table1}, demonstrate that our proposed PRECTR-V2 attains the best performance on both AUC and GAUC metrics, achieving scores of $0.7674$ and $0.6933$, respectively. Compared with our previous work PRECTR, PRECTR-V2 delivers absolute improvements of $0.0093$ and $0.0041$ in AUC and GAUC, significantly outperforming all other baseline methods. These results confirm the effectiveness and superiority of our approach.

\textbf{(2) How is the ablation study results of each component?} 
PRECTR-V2 conotains three core components: (1) Cold-Start Personalized Relevance
Preferences Mining Module,(2) Relevance Exposure Debias Module, and (3) LLM-Distilled CTR-Aligned Lightweight Encoder.
An ablation study was conducted by individually removing each key component to evaluate its contribution to the overall framework. The results, summarized in Table \ref{table2}, show that the removal of any component in PRECTR‑V2 leads to a clear degradation in overall performance. Specifically, Cold‑Start Relevance Preference Mining provides the most substantial performance gain; without it, the relative improvements (RI) in AUC and GAUC drop to only 1.08\% and 0.85\%, respectively. The other two components also contribute positively, as their removal similarly reduces model performance. These findings validate the effectiveness of each proposed enhancement.

\textbf{(3) How is the online performance of PRECTR-V2?} 
We deployed PRECTR-V2 online in Xianyu's search system to evaluate its real-world effectiveness. Using an A/B testing framework, users were randomly assigned to control and experimental groups via MD5 hashing of device IDs to ensure unbiased traffic allocation. The experimental group, receiving over 20\% of total traffic, consistently outperformed the baseline. Specifically, it achieved a 1.39\% lift in per capita orders and a 3.18\% increase in GMV, demonstrating clear business impact. As a result, PRECTR-V2 has now been fully deployed across Xianyu’s search system.

\textbf{(4) How is the PCOC (Predict Click Over Click) metric of PRECTR-V2?} 
In performing relevance exposure debiasing, we employ a pairwise loss to optimize the relative ranking between $\boldsymbol{x}{+}$ and $\boldsymbol{x}{-}$. However, the introduction of such losses inevitably impacts the model’s absolute prediction accuracy, specifically its PCOC (Predict Click Over Click).
To solve this problem, we further introduce a relative order critical distance penalty and a dynamic truncation weighting strategy to strike a balance.
To evaluate its effectiveness, we compute the average PCOC deviation — i.e., the difference from the ideal prediction value of 1.0 — for both PRECTR‑V2 and the baseline model. The results show average deviations of 1.7\% for PRECTR‑V2 and 2.3\% for the baseline.
Although the absolute prediction accuracy of PRECTR-V2 shows a slight decrease, the difference from the baseline remains marginal, indicating that acceptable prediction accuracy has been largely maintained.

\textbf{(5) How does PRECTR-V2 influence the search relevance?} 
Compared to PRECTR, PRECTR‑V2 introduces refined search relevance modeling. To assess its impact, we measured the rate of irrelevant products among the top‑10 results across sampled queries through manual search relevance evaluation.
Results show that PRECTR‑V2 exhibits a 1.09\% relative increase (0.15\% absolute) in irrelevant‑item badcase rate. This minor shift stems from the personalization aspects of relevance estimation and has a negligible impact on overall search relevance performance.

\vspace{-10pt}

\subsection{Hyperparameter Tuning}
\begin{figure}
    \centering
    \includegraphics[width=0.91\hsize,height=0.61\hsize]{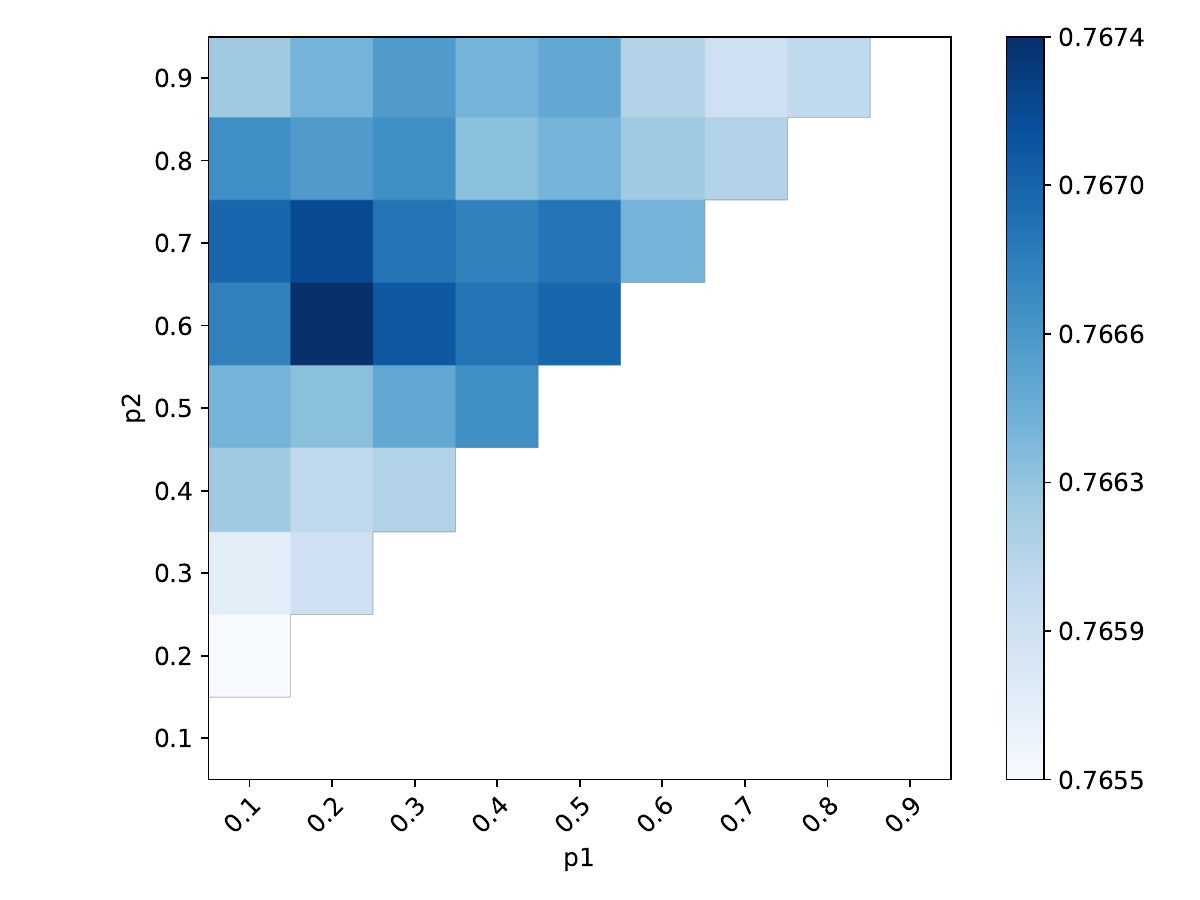}
    \caption{The selection of hyperparameters.}
    \label{hotmap}
\end{figure}

In this subsection, we identify the optimal value of hyperparameters $p1$ and $p2$ ($p1<p2$) in Eq(\ref{eq10}) through two-dimensional grid search.
To be specific, we conduct the search at intervals of 0.1, and traverse the values of $p1$.
Based on the combinations of the values of $p1$ and $p2$, we calculate its corresponding AUC values.
The results are shown in Fig.\ref{hotmap}.
When the $p1$ and $p2$ are set as $0.2$ and $0.6$, the overall model achieves the best performance. Therefore,
we set $p1=0.2$ and $p2=0.6$ in all the experiments.

\section{Conclusion}
\label{conclusion}
This work presents PRECTR-V2, an enhanced unified framework for jointly optimizing search relevance matching and CTR prediction. The proposed approach addresses three critical limitations of its predecessor: ineffective personalization for low-activity users through cross-user relevance preference mining, exposure bias correction via synthetic hard negative sampling and calibrated pairwise ranking, and architectural misalignment via a lightweight, trainable encoder distilled from LLMs. Extensive experiments validate the effectiveness and superiority of PRECTR-V2, demonstrating significant improvements in both relevance understanding and click prediction, ultimately advancing the robustness and practicality of unified search ranking in real‑world applications.

\bibliographystyle{ACM-Reference-Format}
\balance
\bibliography{sample-base}

\end{document}